\documentclass[fleqn,usenatbib]{mnras}

\usepackage{newtxtext,newtxmath}
\usepackage[T1]{fontenc}
\DeclareRobustCommand{\VAN}[3]{#2}
\let\VANthebibliography\thebibliography
\def\thebibliography{\DeclareRobustCommand{\VAN}[3]{##3}\VANthebibliography}

\usepackage{graphicx}	
\usepackage{amsmath}	

\newcommand{\trise}{$t_{\rm IR,rise}$}
\newcommand{\tpeak}{$t_{\rm IR,peak}$}

\title[TDE IR counterparts from dust rings]{The properties of tidal disruption event infrared counterparts produced by dust rings and inference of the observing angle}

\author[R. A. J. Eyles-Ferris]{
Rob A. J. Eyles-Ferris$^{1}$\thanks{E-mail: raje1@leicester.ac.uk}
\\
$^{1}$School of Physics and Astronomy, University of Leicester, University Road, Leicester, LE1 7RH, UK\\
}

\date{Accepted XXX. Received YYY; in original form ZZZ}

\pubyear{\the\year{}}

\begin{document}
\label{firstpage}
\pagerange{\pageref{firstpage}--\pageref{lastpage}}
\maketitle

\begin{abstract}
A substantial fraction of tidal disruption events (TDEs), resulting from a black holes's disruption and accretion of a star, exhibit infrared (IR) counterparts thought to arise from spherical shells of dust reprocessing the TDE emission. Some modelling of TDEs also predicts an angular dependence in their observed properties with more X-ray emission on-axis and more optical emission at higher angles. However, there is growing evidence that X-ray rich TDEs are more likely to exhibit IR counterparts, contradicting the spherical shell model that predicts no significant variation in IR luminosity with angle. Here, I demonstrate that this result naturally follows for dust arranged in a ring instead of a spherical shell. I present a toy model of this scenario and show that on-axis angles result in a brighter counterpart. I also show that on-axis angles result in a delay in the initial rise and off-axis angles may display a double-peaked structure. Crucially, this model also allows the observing angle of the TDE to be constrained. Finally, I demonstrate that this model reproduces the properties of two IR counterparts, including constraining their observing angles and independently inferring an optical plateau, and briefly comment on its application to quasi-periodic eruption counterparts.
\end{abstract}




\section{Introduction}

Tidal disruption events (TDEs) are caused by the destruction and accretion of stars by massive black holes, often the supermassive black holes residing in the nuclei of galaxies. Their observed properties are extremely diverse with significant differences between individual events \citep[e.g.][]{Gezari21}. Assuming that the current sample draws on the same underlying distribution, there are likely to be two factors that drive these differences. First, there is significant evidence that TDE evolution goes in stages and their brightest period is dominated by some form of optically bright compact envelope \citep[e.g.][]{Coughlin14,Metzger22,Sarin24,Tuna25,Mummery25b}. As this envelope cools, it transitions towards a thin disk responsible for the soft X-ray emission. Another possibility is that there is a significant dependence on the observing angle in terms of which emission components. For example, the model of \citet{Dai18}, further developed by \citet{Thomsen22}, indicates that hat observing a TDE closer to the spin axis of the black hole increases X-ray richness while an off-axis angle results in optical emission dominating. While evidence is building that the staged evolution of a TDE dominates, it is likely that both of these factors have an impact on the observed properties of the TDE. For instance, even in the cooling envelope model, the presence of chimneys \citep{Tuna25} may allow X-ray emission from the slowly forming inner disk to `leak' out. Soft X-ray emission is also expected in the similarly quasi-spherical ZEBRA model for observers close to finite density funnels at the axes of the black hole (private comm.). Constraining the observing angle of observed events is crucial for understanding the origins of these components and therefore the underlying structure and evolution of TDEs.

TDEs can also drive other transient emission and a growing number of events exhibit bright infrared (IR) counterparts \citep[e.g.][and see \citet{vanVelzen21b} for a review]{Jiang16,vanVelzen16,Masterson24,Langis25,Grotova25,EylesFerris25}. These counterparts have been linked to circumnuclear dust heated by emission from the TDE and are long lived sources that rise to a peak often hundreds of days after the TDE itself. Typically, this dust is assumed to be arranged in a spherical shell around the black hole \citep[e.g.][]{Lu16,Masterson24} which predicts a fast initial rise to a plateau followed by a long power law-like decay. Most of these counterparts were originally detected by the \textit{Wide-field Infrared Survey Explorer} \citep[\textit{WISE},][]{Wright10}, later reactivated as \textit{NEOWISE} \citep{Mainzer11}, which performed an all-sky survey every six months from 2009 to 2024\footnote{With a pause in operations from 2011 to 2013.}. These long term light curves captured by \textit{WISE} are often found to be roughly consistent with predictions from spherical dust shell models. 

Recent investigation has, however, suggested X-ray rich TDEs i.e. those that may be most likely to be viewed near the axis, are more likely have bright IR counterparts \citep{Langis25,EylesFerris25}. This is difficult to reconcile with a spherical shell model which predicts the IR counterpart should appear essentially the same regardless of the observing angle. In this Letter, I examine the possibility that this discrepancy arises due to the fact the dust is arranged not in a spherical shell but in a ring around the nucleus. Such a geometry is more consistent with the geometrically thick tori inferred in active galactic nuclei \citep[AGN, e.g.][]{Antonucci93}, adding evidence to a connection between these galaxies and TDE hosts. In fact, it may be that TDE hosts are essentially in a `post-AGN' state where active accretion has ceased and the black hole has returned to quiescence. There is some evidence that historic AGN feedback and its cessation may also drive enhanced star formation towards the centre of their host galaxies \citep[e.g.][]{Zubovas17,Su25}, thereby accounting for the observed central concentration of TDE hosts \citep[e.g.][]{Hammerstein21,Newsome25} without requiring a merger to be invoked \citep{Chang26}. This dying burst of star formation could also provide an enhancement to the disruption rate itself. A similar model of a hollowed torus remnant from an AGN has also been suggested to explain AT 2019qiz's unusual IR counterpart \citep{Wu25} but here I examine the case of a surviving ring or torus and develop a toy model of their counterparts. In Section \ref{sec:model}, I present this model, including the underlying assumptions, and in Section \ref{sec:application}, I apply it to two IR counterparts. I briefly examine the predictions for the TDE-related phenomena of quasi-periodic eruptions (QPEs) in Section \ref{sec:qpe_counterparts} and finally summarise my conclusions in Section \ref{sec:conclusions}.

\section{A dust ring model for TDE IR counterparts}
\label{sec:model}

In this Section, I present a model of an IR counterpart produced from a ring of dust surrounding the TDE. I emphasise that this is a toy model and aspects such as the composition of the dust, its temperature evolution or internal scattering of the IR emission is not examined here. I also assume the dust lies entirely in a ring (thin or thick). It is of course also possible that the distribution is broadly spherical but with significantly more dust concentrated in the ring\footnote{This would reduce the rise times of on-axis events but the impact is likely to be small with the ring dominating the morphology of the light curves.}.

\subsection{A TDE emission model}
\label{sec:tde_emission}

To derive the properties of an IR counterpart, the emission of the TDE itself needs to be defined. This will then be fed into a reprocessing model to calculate the light curve. It should also be noticed that the specific SED of the TDE is not examined in detail here. The SED of TDE emission is complex and the underlying components and their evolution are not fully understood. However, from observational evidence, it is likely that the IR counterpart is more closely tied to the optical evolution than other components. This may, at first glance, be somewhat surprising due to the intrinsically hotter radiation field of an X-ray dominated TDE which results in a greater fraction of the emission being absorbed to be reprocessed by the dust. For instance, under the model of \citet{Draine03} with a fixed dust concentration, $\sim1.5$ times as much radiation from a 60 eV blackbody will be absorbed as from a $2\times10^4$ K blackbody. However, the optical component of a TDE tends to be significantly, typically an order of magnitude or more, more luminous than the X-ray as aptly demonstrated by Figure 8 of \citet{Hammerstein23}. While this could, of course, be due to the observing angle\footnote{In this situation, the evolutionary stage does not have a significant impact as this is relative to the optical and presumably bolometric peak.}, it does imply that the vast majority of emission absorbed by the dust comes from the optical component. This is further borne out by the rise times observed in IR counterparts. For instance, 1eRASS J1436 from the sample of \citet{EylesFerris25} was found to be consistent with an intrinsic peak at least 40 days before the earliest possible X-ray peak\footnote{I note that this was derived using the spherical shell model but this particular result should be similar in both the spherical shell and dust ring models.}.

I therefore adopt the Gaussian rise, power law decay model that \citet[][see also \citet{Hammerstein23}]{vanVelzen21} show to accurately match the optical luminosity evolution of TDEs and assume a fixed temperature throughout. The intrinsic luminosity of the TDE is given by
\begin{equation}
    L_{\rm int}(t) = L_{\rm pk}\times
    \begin{cases}
        e^{-(t-t_{\rm pk})^2 / 2t_{\rm r}^2}, & t \leq t_{\rm pk},\\
        \left((t - t_{\rm pk} + t_{\rm fb}) / t_{\rm fb}\right)^{-\frac{5}{3}}, & t > t_{\rm pk},
    \end{cases}
    \label{eq:Lint}
\end{equation}
where $t_{\rm pk}$ is the peak time of the TDE, and $t_{\rm r}$ and $t_{\rm fb}$ are rise and fallback timescales respectively. These are fixed to $t_{\rm r} = 10$ days and $t_{\rm fb} = 100$ days in the modelling below based on the results of \citet{Hammerstein23}. This simple model is a straightforward representation of TDE emission. However, features such as a late time plateau may also need to be incorporated to fully capture observed behaviour in some TDEs. This is explored in Section \ref{sec:application} in the context of AT 2019dsg.

\subsection{A thin ring}

\begin{figure}
    \centering
    \includegraphics[width=\columnwidth]{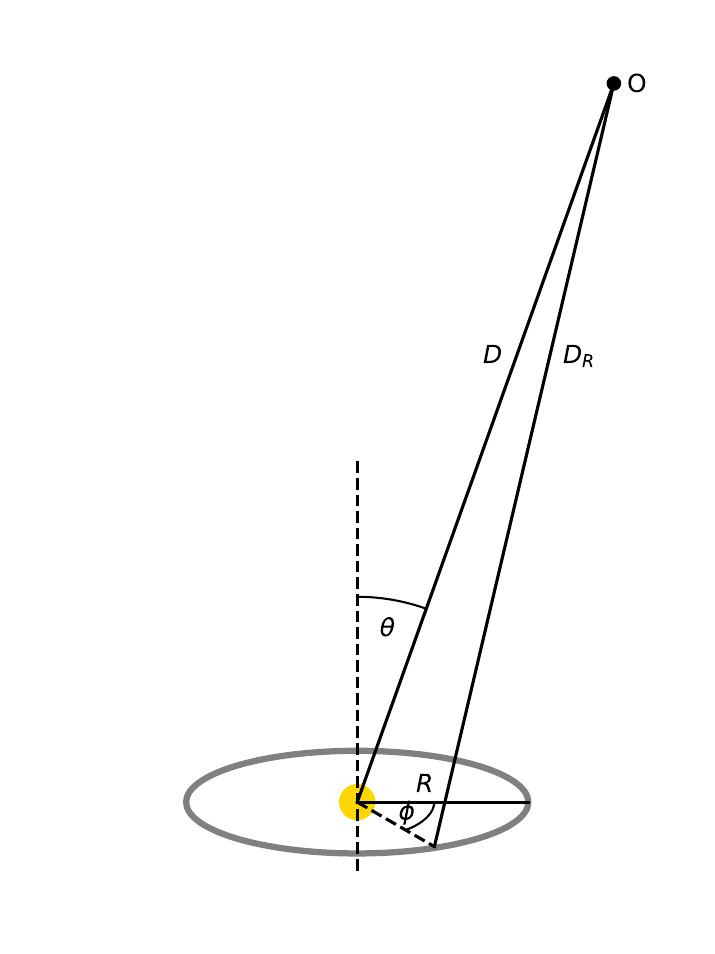}
    \caption{The geometry of the TDE-dust ring system. The observer is positioned at O, a distance $D$ from the TDE (yellow) and $D_R$ from the point on the dust ring (grey) defined by angle $\phi$.}
    \label{fig:geometry}
\end{figure}

I show the assumed geometry of the TDE-dust ring system in Figure \ref{fig:geometry}. The TDE is positioned in the centre of a thin ring of dust of radius $R$. The observer is positioned at O at a distance $D$ from the TDE with an observation angle of $\theta$ relative to the axis of the system. Each point on the ring is defined by an angle $\phi$ with the closest point on the ring to the observer set to $\phi = 0$. 

It is trivial to show the path length, $D_R$, from a given point on the ring is given by
\begin{equation}
    D_R(\phi) = \left(D^2 + R^2 - 2DR\sin(\theta)\cos(\phi)\right)^{\frac{1}{2}}.
    \label{eq:path_length}
\end{equation}
By subtracting the path length of TDE light from the time it reaches the dust ($D-R$), the delayed time relative to the TDE itself, $\tau$, can then be calculated as a function of $\phi$. The reprocessed emission observed from a given arc of length $Rd\phi$ will be $Rd\phi\times\epsilon L_{\rm int}(\tau)/4\pi R^2$ where $\epsilon$ is the efficiency (assumed to be 1 for simplicity here). To find the total observed luminosity of the thin ring at a given time, $t$, I then integrate over $\phi$ i.e.
\begin{equation}
    L_{\rm IR,thin}(t) = \frac{1}{4\pi R}\int_0^{2\pi} L_{\rm int}(\tau(\phi)) d\phi.
    \label{eq:L_thin}
\end{equation}

\begin{figure}
    \centering
    \includegraphics[width=\columnwidth]{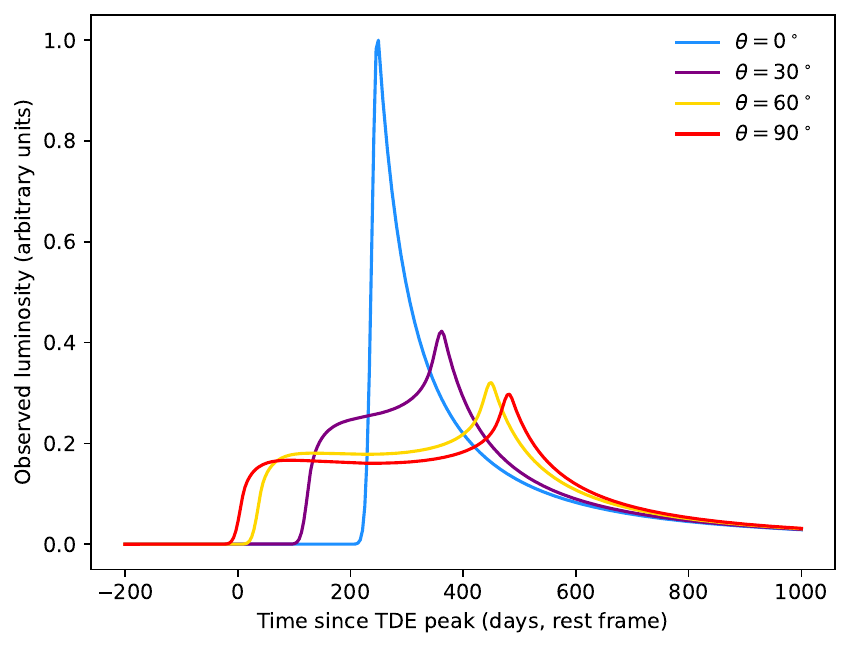}
    \caption{The IR light curve of a TDE-dust ring system observed from different observing angles assuming a thin ring.}
    \label{fig:thin_lcs}
\end{figure}

I assume $R=0.2$ pc, roughly consistent with the mean of the distribution identified by \citet{Masterson24}\footnote{While \citet{Masterson24} assume a spherical model rather than the ring investigated here, the rise timescale for a given dust radius should be broadly comparable, particularly in off-axis cases.} and $D = 475.8$ Mpc, corresponding to the luminosity distance at $z=0.1$ in a Planck cosmology \citep{Planck18}. Figure \ref{fig:thin_lcs} shows the resulting light curves for increasingly off-axis angles.

It is immediately clear that on-axis counterparts are significantly brighter by a factor of a few than those at higher angles. As X-ray bright TDEs are viewed more on-axis, their IR counterparts will be accordingly easier to detect. A ring model therefore accounts for the larger fraction of sources in this sample with IR counterparts compared to TDES with faint or no X-ray emission.

Figure \ref{fig:thin_lcs} also demonstrates that the light curves produced from a ring have morphologies that vary significantly with observing angle and are also substantially different from those predicted in a spherical shell model. While the fully on-axis case rises rapidly and fades as a power law, essentially a true reflection of the TDE as $\tau$ is constant across the ring, moving off-axis results in a light curve that rapidly rises to a plateau or decay phase. Similar plateauing behaviour is predicted from a spherical shell but for a ring, there is a final rise to peak prior to the final decay. A similar second rise or even second peak has been seen in a small minority of observed events such as WTP 14acnjbu from the sample of \citet{Masterson24}. Another key prediction from this modelling is that there will be a delay between the rises of the TDE and the infrared counterpart that increases for smaller observing angles. This is not expected in a spherical shell as dust will also be in the line of sight to the TDE hence the TDE and its counterpart will rise simultaneously. However, the sparsely sampled \textit{WISE} light curves that currently dominate the IR counterpart sample mean that identification of these features is extremely difficult. Wider confirmation of late peaks and delayed rises will be able to differentiate between the spherical shell and ring geometries.

Returning to Equation \ref{eq:path_length}, it can be deduced that the final peak occurs approximately when peak emission from the far side of the ring (i.e. $\phi = \pi$) has reached the observer\footnote{This also depends slightly on the TDE model assumed and, for instance, a slow fading example may peak later than predicted by Equation \ref{eq:t_IR_peak}.} while the light curve starts to rise when light from the near side ($\phi = 0$) first reaches the observer. Setting $\phi$ to these values in Equation \ref{eq:path_length}, I derive
\begin{equation}
    t_{\rm IR,rise} = \frac{\left(D^2+R^2-2DR\sin(\theta)\right)^{\frac{1}{2}} - (D-R)}{c}
    \label{eq:t_IR_rise}
\end{equation}
and
\begin{equation}
    t_{\rm IR,peak} = \frac{\left(D^2+R^2+2DR\sin(\theta)\right)^{\frac{1}{2}} - (D-R)}{c}
    \label{eq:t_IR_peak}
\end{equation}
where \trise~and \tpeak~are the times relative to these phases in the light curve of the TDE itself.

These two equations reveal an important result. While $\theta$ and $R$ initially appear degenerate, if the light curve is sufficiently constrained, there is only a small region of $\theta$-$R$ parameter space that can satisfy both Equations \ref{eq:t_IR_rise} and \ref{eq:t_IR_peak}. In other words, if the dust is indeed in a thin ring, the IR light curve allows not only the dust radius, but the observing angle of the TDE to be directly inferred. However, as mentioned above, constraining the IR light curve is not trivial
. 
High cadence, sensitive observations over a long time period, for instance, through dedicated ground based campaigns or through a nearby IR bright TDE lying within one of the \textit{Nancy Grace Roman Space Telescope}'s High-Latitude Time-Domain Survey\footnote{\url{https://roman-docs.stsci.edu/roman-community-defined-surveys/high-latitude-time-domain-survey}} fields, will be necessary to tightly constrain the dust geometry and observing angle.

\subsection{Other geometries and attenuation}
\label{sec:attenuation}

It is unlikely, however, that circumnuclear dust lies in a single thin ring. In the following Sections, I therefore also explore two other possible geometries - a radially thick ring\footnote{\citet{Wu25} essentially explore the vertically thick and radially thin case.} and a thick torus. These geometries can be modelled as a superposition of thin rings. However, an additional effect should be also considered, the attenuation of light from the TDE as it travels through inner regions of the dust.

To address the attenuation, I introduce an attenuation coefficient, 
\begin{equation}
    A(d_A) = \left(\frac{d_A+d_c}{d_c}\right)^{-\alpha},  
\end{equation}
where $d_A$ is the distance travelled in attenuating material (the thick ring or torus), $d_c$ is a characteristic distance taken to be 0.01 pc and I set $\alpha=1/2$ (chosen for visibility in Figures \ref{fig:thick_lcs} and \ref{fig:torus_lcs}). While this is very much a simplification of the situation and it will likely vary somewhat between TDEs depending on the number density and elemental composition of the rings, the full radiative transfer simulations required are beyond the scope of this work and this provides a reasonable approximation.

The IR emission will also, of course, be attenuated as it passes through the dust. It may also be internally scattered resulting in a delay depending on the line of sight. While these factors are likely to be non-negligible, they should still make a relatively small impact and therefore I neglect them in this toy model.

\subsection{A radially thick ring}

\begin{figure}
    \centering
    \includegraphics[width=\columnwidth]{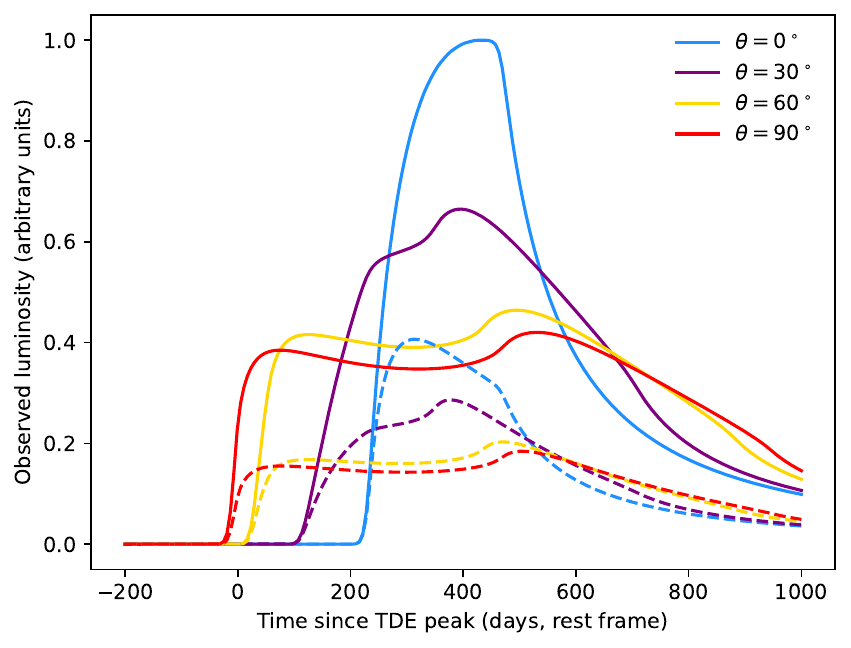}
    \caption{Mid-infrared light curves from a radially thick ring when the TDE light is unattenuated (solid) and attenuated (dashed).}
    \label{fig:thick_lcs}
\end{figure}

The setup for a radially thick ring is a simple superposition of thin rings with increasing radius and increasingly attenuated $L_{\rm int}$. In this case, I take an inner radius of 0.2 pc and an outer radius of 0.4 pc. To derive the light curve from this radially thick ring, I integrate over the individual thin rings as
\begin{equation}
    L_{\rm IR,thick}(t) = \int^{R_{\rm outer}}_{R_{\rm inner}} A(R)~L_{\rm IR,thin}(t, R)~dR.
    \label{L_thick}
\end{equation}
The resulting light curves, both unattenuated and attenuated, are shown in Figure \ref{fig:thick_lcs}.

The characteristic shape of the single thin ring is still present in these light curves i.e. a hump rising to a sharp peak before an approximately power law decline. However, due to the multiple rings, this is significantly smoothed, particularly around peak time, even in the attenuated light curves dominated by the inner rings. In the unattenuated case, there is also more significant decay following the initial rise at highly off-axis angles. Again, an on-axis angle results in a significantly higher peak flux. However, the difference is somewhat smaller than in the thing ring case. For instance, here the ratio of peak flux for an observing angle of 0\degr~to that for 30\degr~is $\sim1.5$ (for both unattenuated and attenuated) as opposed to $\sim2.5$ for the thin ring.

Finally, the peak is significantly broadened by the contribution across the thick ring. This means that while Equation \ref{eq:t_IR_rise} holds for $R=R_{\rm inner}$, \ref{eq:t_IR_peak} no longer holds for the same radius. However, the observing angle can still be reasonably constrained through light curve fitting.

\subsection{A thick torus}

\begin{figure}
    \centering
    \includegraphics[width=\columnwidth]{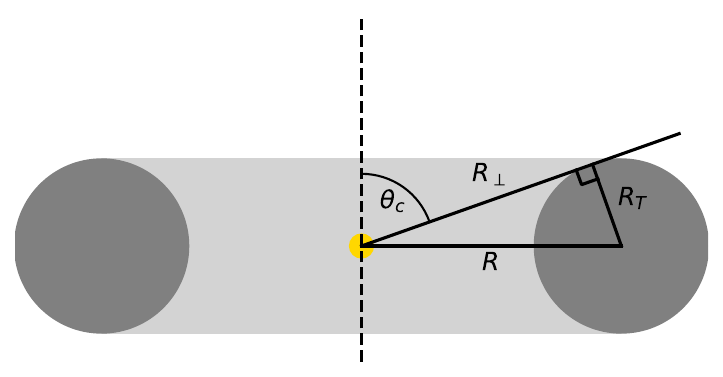}
    \caption{A cross section of through the torus model showing the TDE (yellow) and the dusty torus (grey). The tangent and critical angle, $\theta_c$, are also shown (see text for details).}
    \label{fig:torus_geometry}
\end{figure}

The final model I consider here is a thick torus. I assume a simple geometry for the torus consisting of a ring with a radius of $R$ and a circular cross section of radius $R_T$. I show a cross section of this geometry in Figure \ref{fig:torus_geometry}. The torus is again a superposition of thin rings, attenuated according to their position in the torus relative to the TDE, and I integrate both radially and vertically over these to derive the final light curves. An immediate consideration is the impact of the new vertical thickness in this model, specifically how it affects \trise.

By considering the tangent from the TDE to the dust torus, it is clear that for observing angles above a critical angle, $\theta_c=\arccos\left(R_T / R\right)$, the line of sight from the observer to the TDE passes through the torus. This means that reprocessed emission from the torus will arrive at the same time as light from the TDE i.e. \trise$\sim 0$, although it should be noted that the rise may initially be very slow due to the relative paucity of dust on that particular line of sight. Even for more on-axis angles, the path length will be substantially shortened hence \trise~may be significantly reduced compared to a thin or radially thick ring of similar radius. For the torus, Equation \ref{eq:t_IR_rise} becomes
\begin{equation}
    t_{\rm IR,rise,torus} = \begin{cases}
        0, & \theta > \theta_c \\
        R_\perp+(D^2+R_\perp^2 \\ \hspace{0.5cm} -2DR_\perp\cos(\theta_c-\theta))^{\frac{1}{2}} - D)/c, & \theta \leq \theta_c.
    \end{cases}
    \label{eq:t_IR_rise_torus}
\end{equation}
where $R_\perp = \left(R^2 - R_T^2\right)^{1/2}$ is the distance from the TDE to the tangent point. It should also be noted that if $R_T\ll R$, the resulting light curves start to converge towards the thin ring case discussed above (albeit smoother). In these cases, $R_\perp$ and $\theta_c$ will tend towards $R$ and 90\degr~respectively and Equation \ref{eq:t_IR_rise_torus} will tend towards Equation \ref{eq:t_IR_rise}.

\begin{figure}
    \centering
    \includegraphics[width=\columnwidth]{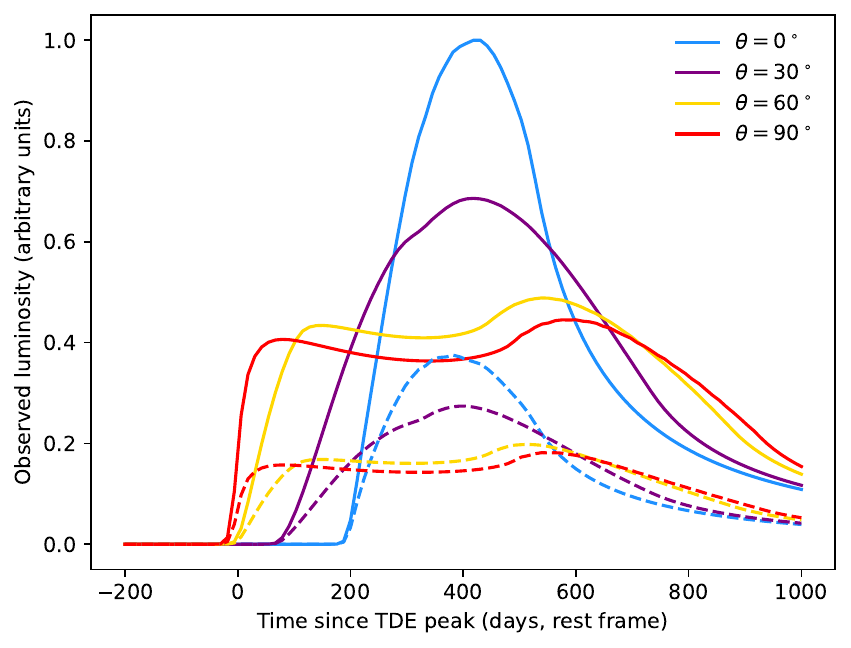}
    \caption{Mid-infrared light curves from a torus when the TDE light is unattenuated (solid) and attenuated (dashed).}
    \label{fig:torus_lcs}
\end{figure}

In Figure \ref{fig:torus_lcs}, I show the unattenuated and attenuated light curves when taking $R=0.3$ pc and $R_T=0.1$ pc with the same TDE setup as previously used. The light curves produced by a torus strongly resemble those of a thick disk but with the sharper features further smoothed. This can be particularly seen with the 30\degr~observing angle where both the unattenuated and attenuated light curves display an almost linear rise rather than any form of plateau. Again, however, qualitatively similar light curves can be seen in \citet{Masterson24}'s sample, particularly WTP 14adbwvs. Depending on the attenuation assumed, there may also still be a somewhat double-peaked appearance at higher angles but it is likely that many of these features will be smoothed out.

Equation \ref{eq:t_IR_rise_torus} provides an accurate prediction for \trise. Similarly to the thick ring case, it is difficult to precisely predict the peak time when $R_T$ is a substantial fraction of $R$. However, for $R \gtrsim 10R_T$ Equation \ref{eq:t_IR_peak} remains a good indicator. In both cases, as shown in Section \ref{sec:application}, light curve fitting is effective for constraining the properties of the dust.

\section{Applying the dust ring model to TDE counterparts}
\label{sec:application}

\begin{table*}
    \caption{The priors for the torus model fits to the example sources' light curves.}
    \label{tab:priors}
    \centering
    \begin{tabular}{cccc}
    \hline
    Parameter & 1eRASS J075803.3+075526 priors & \multicolumn{2}{c}{AT 2019dsg priors} \\
    & & Initial & Plateau \\\hline
    $R$ (pc) & $0.05 < R < 3.5$ & $0.05 < R < 3.5$ & $0.01 < R < 0.5$ \\
    $R_T$ (pc) & $R_T < R$ & $R_T < R$ & $R_T < R$ \\
    $\theta$ & 0\degr$\leq \theta \leq90$\degr &  0\degr$\leq \theta \leq90$\degr &  0\degr$\leq \theta \leq90$\degr \\
    $t_{\rm pk}$ (MJD) & $57483 < t_{\rm pk} < 60500$ & 58605.38 & 58605.38 \\
    $t_{\rm plat}$ (MJD) & --- & --- & $58780.38 < t_{\rm plat} < 59205.38$ \\
    \hline
    \end{tabular}
\end{table*}

\begin{table*}
    \caption{The posterior distributions of the parameters fitted for each of our example sources. Note that for AT 2019dsg, $t_{\rm pk}$ is not fitted but is taken from \citet{Hammerstein23}.}
    \centering
    \renewcommand{\arraystretch}{1.4}
    \begin{tabular}{cccccccc}
    \hline
     Source & $z$ & $D$ (Mpc) & $R$ (pc) & $R_T$ (pc) & $\theta$ (\degr) & $t_{\rm pk}$ (MJD) & $t_{\rm plat}$ (MJD) \\
    \hline
    1eRASS J075803.3+075526 & 0.0955$^1$ & 453.1 & $2.43\pm0.44$ & $2.06\pm0.47$ & $58^{+14}_{-9}$ & $57530\pm40$ & --- \\
    AT 2019dsg (initial) & 0.051$^2$ & 234.6 & $2.75^{+0.53}_{-0.77}$ & $0.31\pm0.20$ & $87\pm2$ & 58605.38 & --- \\
    AT 2019dsg (plateau) & 0.051$^2$ & 234.6 & $0.26\pm0.07$ & $0.23\pm0.09$ & $74^{+12}_{-36}$ & 58605.38 & $58907\pm42$ \\
    \hline
    \multicolumn{7}{l}{$^1$\citet{Alam15}; $^2$\citet{Hammerstein23}}
    \end{tabular}
    \label{tab:posteriors}
\end{table*}

\begin{figure*}
    \centering
    \includegraphics[width=0.66\columnwidth]{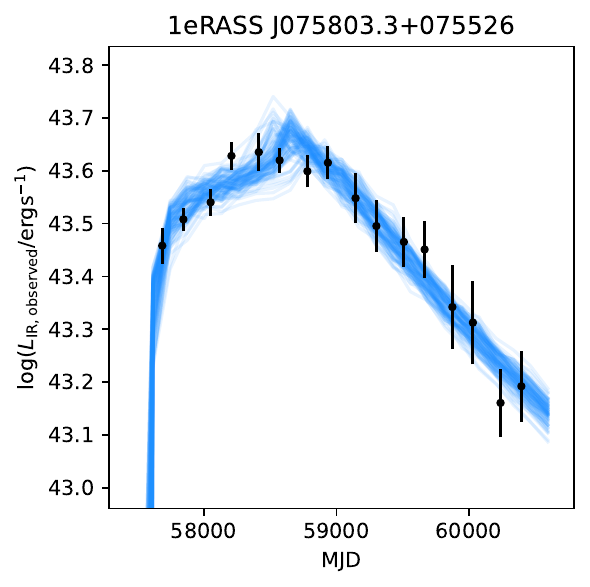}
    \includegraphics[width=0.66\columnwidth]{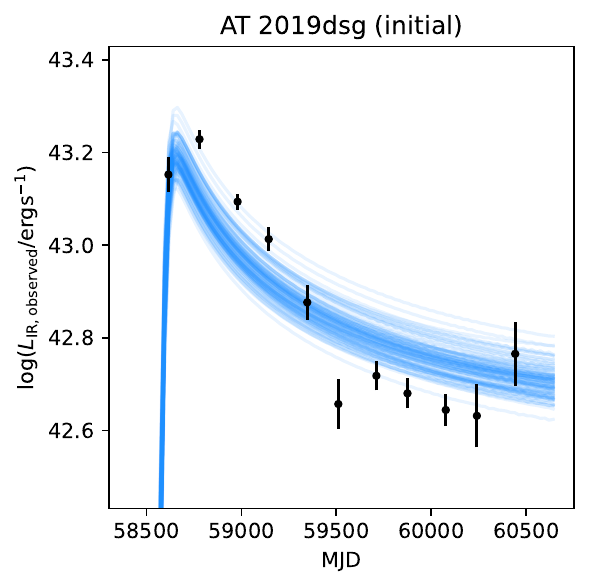}
    \includegraphics[width=0.66\columnwidth]{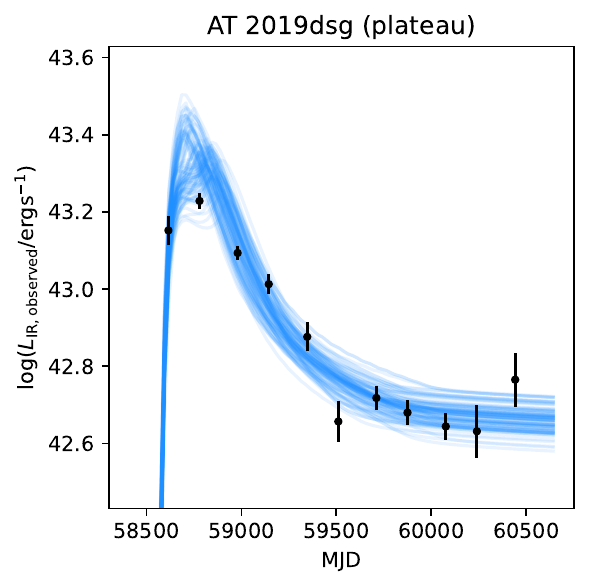}
    \caption{1eRASS J075803.3+075526 (left) and AT 2019dsg (initial model in centre and model modified with an optical plateau on right, see text for details) fitted with the torus light curve model. For each source, 100 random 1-$\sigma$ traces from the MCMC posterior distribution are shown (blue). Note the short term variability seen in 1eRASS J075803.3+075526's light curve is a numerical artefact that does not have a significant effect on the results.}
    \label{fig:fitted_lcs}
\end{figure*}

To test the dust ring model, I now apply it to two examples of observed IR counterparts. These consist of 1eRASS J075803.3+075526 from the sample of \citet{EylesFerris25}, and the well studied TDE AT 2019dsg \citep[e.g.][]{Lee20,Cendes21,Matsumoto22,Hammerstein23}.

I obtained host subtracted \textit{WISE} photometry using the method described in \citet{EylesFerris25}. To briefly review this process, individual \textit{W1} and \textit{W2} band images were obtained using the Simple Image Access v2\footnote{\url{https://irsa.ipac.caltech.edu/ibe/sia.html}} service and stacked using \texttt{astroalign v2.6.1}. Template AllWISE images were obtained from the unWISE cutout service \citep{Lang14,Meisner17a,Meisner17b} and subtracted using \texttt{PyZOGY} \citep{Zackay16,Guevel21}. The residual flux was measured using PSF photometry and calibrated against the AllWISE Source Catalogue \citep{Cutri13}. For each epoch where a residual was detected to at least a 3-$\sigma$ significance in both bands, a blackbody model was fitted using \texttt{astropy v7.1.0} \citep{astropy13,astropy18,astropy22} and used to derive the observed luminosity light curve.

Here, I use only the torus model to fit the light curves and initially assume the model in Equation \ref{eq:Lint} for the TDE emission itself. For AT 2019dsg, I modify this to use the values derived by \citet{Hammerstein23} in their fit to the optical light curve i.e. $t_{\rm pk} = {\rm MJD}~58605.38$, $t_{\rm r} = 23$ days and $t_{\rm fb} = 72$ days. I also set a lower bound of $t_{\rm pk} > {\rm MJD}~57483$, the last \textit{WISE} non-detection, for 1eRASS J075803.3+075526's counterpart. I used \texttt{emcee v3.1.6} \citep{ForemanMackey13} sampler for this fitting, using 16 walkers over 5000 iterations of which the first 500 were discarded as burn-in. Six parameters were fitted, $R$, $R_T$, $\theta$, $t_{\rm pk}$\footnote{Note that this is the peak time of the TDE itself rather than its IR counterpart.}, the normalisation and a parameter $f$ which describes underestimates in the observed errors. Based on preliminary tests, the attenuation index was fixed to $\alpha=1$ (i.e. an approximately constant torus density and composition) and for each source, $D$ was fixed to their luminosity distance. The sets of priors assumed for the physical parameters ($R$, $R_T$, $\theta$ and $t_{\rm pk}$) is given in Table \ref{tab:priors}, their posterior distributions are given in Table \ref{tab:posteriors} and the fitted light curves shown in Figure \ref{fig:fitted_lcs}.

As shown in Figure \ref{fig:fitted_lcs}, the torus model does indeed produce an excellent match to the observed behaviour of 1eRASS J075803.3+075526. The dust radius, at first glance, seems significantly larger than that inferred in \citet{EylesFerris25} but the large $R_T$ places dust a few tenths of a parsec from the TDE consistent with the $\sim 0.5$ pc identified there. Importantly, the model has constrained the observing angle to within 10\degr~or so and implies that this event is being observed through the torus itself. The assumed attenuation model gives $A\sim0.004$ which may result in the lack of optical counterpart identified by \citet{EylesFerris25}.

AT 2019dsg is a slightly more complicated proposition primarily due to the late plateau seen in the IR light curve\footnote{This is not unique to the torus model and would also pose an issue for a spherical shell model that assumes the same underlying TDE emission.}. The initial set of priors do result in an adequate fit which places the observed plateau occurs between two peaks with the second peak occurring some time in the mid 2030s. However, due to the rapid rise of the IR counterpart and the large dust radius required, this requires an essentially fully off-axis observing angle. It is arguably more likely that the IR plateau is due to a similar plateau in the TDE itself \citep[e.g.][]{{Mummery24,Mummery25}} and in fact a similar slow fade or plateau can be observed in AT 2019dsg's optical light curve from about 200 days post peak (see Figure 17 in \citet{Hammerstein23}). I therefore modify Equation \ref{eq:Lint} to add a late time plateau component i.e. for $t > t_{\rm plat}$,
\begin{equation}
    L_{\rm int}(t) = L_{\rm pk} \times \left((t_{\rm plat}-t_{\rm pk}+t_{\rm fb})/t_{\rm fb}\right)^{-\frac{5}{3}},
    \label{eq:tplat}
\end{equation}
i.e. the emission is constant after $t_{\rm plat}$. This is consistent with predictions for the emission from a spreading accretion disk as inferred from the widespread presence of UV-plateaus across the TDE sample \citep[e.g.][]{Mummery20,Mummery25}. The fit is repeated with $t_{\rm plat}$ as a free parameter and modified priors as given in Table \ref{tab:priors}. I find that the fast rise time could still imply a relatively off-axis observing angle but a much smaller dust radius is required. In fact, due to the large $R_T/R$ ratio, there is significant dust less than a tenth of a parsec from the TDE consistent with the suggestion that the neutrino production linked to this source is driven by the presence of nearby dust \citep{Winter23}. The large $R_T/R$ ratio also gives $\theta_c \sim 32.2$\degr~suggesting that AT 2019dsg is also being viewed through the optically thin torus. At the inferred $\theta$, $A\sim0.02$ suggesting a substantial fraction of the intrinsic luminosity could remain unobserved. Finally, the time of the inferred plateau is consistent with that seen in the optical light curve. This is the first time that an IR feature has been used to infer the optical behaviour of a TDE. Further analysis of such features could be used to derive the properties of underlying optical plateaus and potentially, through the scaling relations of \citet{Mummery24}, the black hole mass.

In both cases, the observing angle is relatively high given that these are both X-ray rich TDEs. This may be due to the assumed geometry in these cases which requires such an angle to replicate the rapid rise. The viewing angle through the torus may also lead to the X-rays themselves also being absorbed. As discussed in Section \ref{sec:tde_emission}, the fraction of X-rays absorbed is comparable to that of the optical emission and therefore they remain observable. It is also clear that X-rays are being emitted along the line of sight and therefore can be absorbed by the torus which reduces the fraction of optical emission that needs to be absorbed. This may result in a slightly different geometry for the torus and examining this possibility further could aid in understanding the emission geometry of the TDE itself. 

Further aspects that have been neglected in this simple toy model such as internal scattering are also likely to further improve the model and reconcile some of the remaining disparate features. Overall, however, the torus model has proven effective in reproducing the properties of these observed light curves and constraining the viewing angles of these objects.

\section{IR counterparts driven by quasi-periodic eruptions}
\label{sec:qpe_counterparts}

Quasi-periodic eruptions (QPEs) are a phenomenon that can emerge following a TDE. QPEs are repeating spikes of soft X-rays and may be the result of extreme mass ratio inspirals (EMRIs) interacting with the accretion disk formed by a TDE. The emission from QPEs can also drive IR counterparts but with distinct properties compared to those expected from TDE driven sources. \citet{Pasham25} apply a spherical dust shell and show that, unlike TDEs, QPEs would drive a linear rise to a long, oscillating plateau phase and point out this behaviour is consistent with that observed in the TDE/QPE source AT 2019qiz\footnote{Although I note that \citet{Wu25} show that the IR counterpart is also broadly consistent with a TDE source where the dust is arranged in an incomplete torus.}. As it seems unlikely that QPEs and TDEs would have inherently different dust distributions, here, I also apply the dust ring model to QPEs.

\begin{figure*}
    \centering
    \includegraphics[width=0.66\columnwidth]{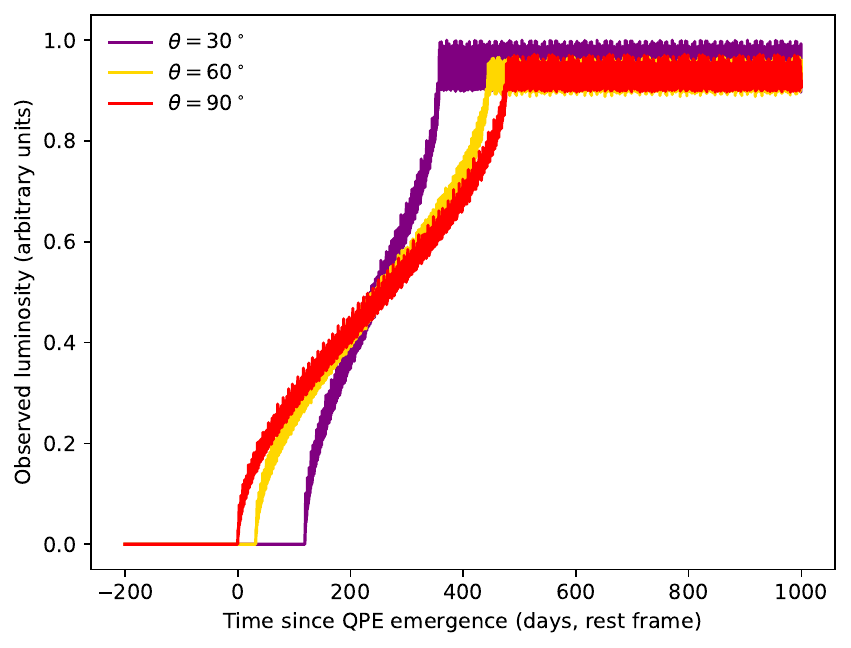}
    \includegraphics[width=0.66\columnwidth]{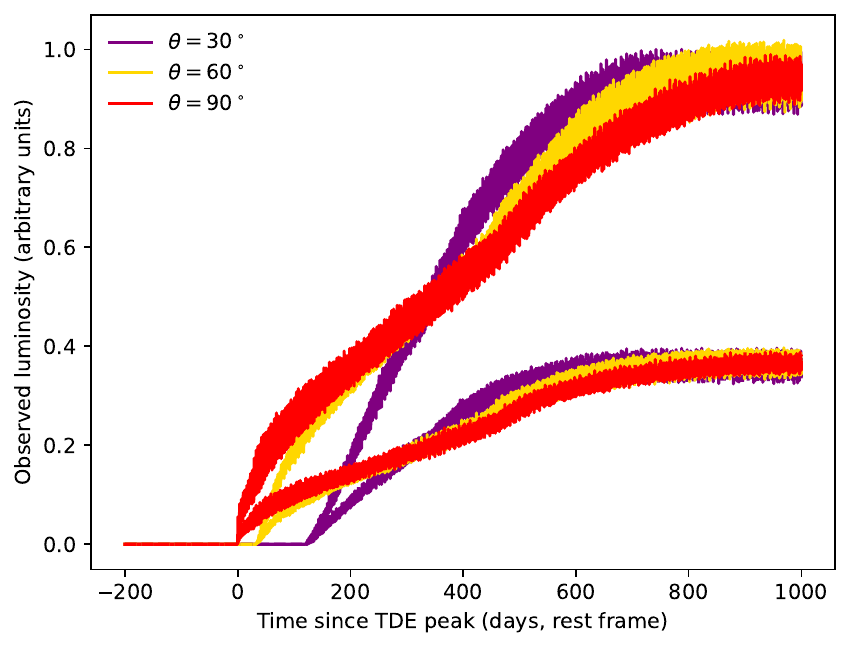}
    \includegraphics[width=0.66\columnwidth]{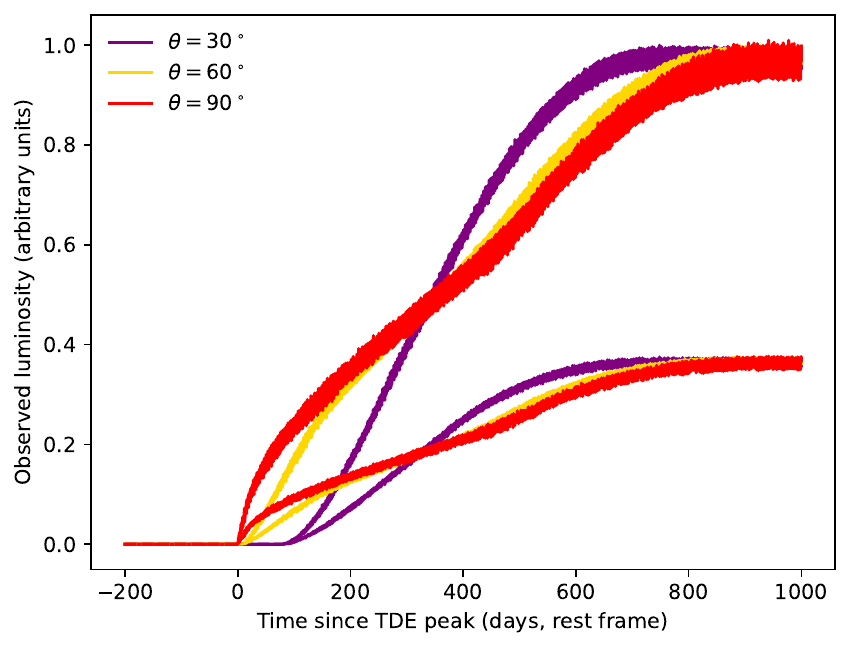}
    \caption{The light curves of QPE-dust ring systems observed from different observing angles assuming \textbf{left} a thin ring, \textbf{middle:} a radially thick ring, and \textbf{right:} a torus. Note the y axis scaling differs between panels. In the thick ring and torus model panels, the lower set of light curves have attenuation applied as described in Section \ref{sec:attenuation}.}
    \label{fig:qpe_lcs}
\end{figure*}

The QPE is modelled as a source that emits at $L_{\rm QPE}$ for 3 ks periodically every 30 ks and has a luminosity of zero at all other times. I use the same three dust geometries as in Section \ref{sec:model} and apply this model of $L_{\rm int}$. The resulting light curves are shown in Figure \ref{fig:qpe_lcs}. The $\theta=0\degr$ light curve is not included to ensure clarity in the plots. However, in the thin ring case, the model predicts a rapidly varying source similar to the QPE itself. In the thick ring and torus cases, a similar rise to a plateau is predicted as in the plots but with significantly larger oscillations.

In all cases, the IR counterpart smoothly rises to an oscillating plateau. While the rise the models predict are not strictly linear, they would likely appear to be so when sampled at a six month cadence as in the case of the \textit{WISE} observations of AT 2019qiz. The much longer rise time predicted in the case of the thick ring and torus models is also consistent with AT 2019qiz which would otherwise appear to have a significantly larger radius than other MIR counterparts if the dust is arranged spherically. In conclusion, it seems the dust model is also consistent with the properties of observed QPE counterparts.

\section{Conclusions}
\label{sec:conclusions}

In this Letter, I have investigated the IR counterparts to TDEs expected from circumnuclear dust arranged in a ring rather than a spherical shell. These models include a thin ring, a radially thick ring and a thick torus. My conclusions are summarised below:
\begin{itemize}
    \item The light curves produced by dust rings show two smoking gun features distinct from those expected from a spherical shell model; a late rise to the final peak and a delayed rise on a timescale inversely correlated with the observing angle. The morphology of the light curves can be used to directly infer both the dust radius \textit{and} the observing angle of the TDE.
    \item More on-axis observing angles result in a significantly brighter (factor a few) IR counterpart accounting for the greater prevalence of counterparts observed in on-axis X-ray bright TDEs suggested by \citet{Langis25} and \citet{EylesFerris25}.
    \item The torus model can reproduce real IR counterparts and constrain their observing angles to 1-$\sigma$ errors of 10 to 20\degr. In the case of AT 2019dsg, an optical plateau was also inferred demonstrating IR counterparts can also be used to examine the intrinsic properties of a TDE.
    \item A broadly smooth and linear rise to a plateau is predicted for QPE counterparts consistent with the observed nature of AT 2019qiz and other models for QPE counterparts \citep{Pasham25,Wu25}.
\end{itemize}
Future improvements to the toy model presented here will include full modelling of the radiative transfer and other effects such as internal scattering of the IR emission. The impact of these is likely to improve the match to observed behaviour and allow the dust properties and observing angle to be better constrained. However, dedicated observing campaigns will be required to fully understand the nature and geometry of the dust. The imminent launch of the \textit{Nancy Grace Roman Space Telescope} and \textit{Near-Earth Object Surveyor} missions offer an opportunity for such studies. Confirming that the dust does indeed lie in rings will allow inference of observing angles across the TDE population, crucial for understanding the nature of their emission and the underlying physics of these extreme accretion events.

\section*{Acknowledgements}

I thank the referee for their invaluable comments which have significantly improved this work, EREF for useful discussion, and the Museo Elder de la Ciencia y la Tecnolog\'{i}a in Las Palmas de Gran Canaria whose black hole `simulator' provided the seed of this idea.

\section*{Data Availability}

The subtracted \textit{WISE} photometry used in Section \ref{sec:application} will be made available in a public database in the future and will be provided on reasonable request to the author in the mean time. 

\bibliographystyle{mnras}
\bibliography{main}



\bsp
\label{lastpage}
\end{document}